\documentstyle[pra,aps,epsfig]{revtex}
\begin{document}
\draft
\title{Real measurements and Quantum Zeno effect}
\author{Julius Ruseckas and B. Kaulakys}
\address{Institute of Theoretical Physics and Astronomy,\\
A. Go\v{s}tauto 12, 2600 Vilnius, Lithuania}
\date{\today}
\maketitle

\begin{abstract}
In 1977, Mishra and Sudarshan showed that an unstable particle would never be
found decayed while it was continuously observed. They called this effect the
quantum Zeno effect (or paradox). Later it was realized that the frequent
measurements could also accelerate the decay (quantum anti-Zeno effect). In
this paper we investigate the quantum Zeno effect using the definite model of
the measurement. We take into account the finite duration and the finite
accuracy of the measurement. A general equation for the jump probability during
the measurement is derived. We find that the measurements can cause inhibition
(quantum Zeno effect) or acceleration (quantum anti-Zeno effect) of the
evolution, depending on the strength of the interaction with the measuring
device and on the properties of the system. However, the evolution cannot be
fully stopped.
\end{abstract}

\section{Introduction}

Theory of measurements has a special status in quantum mechanics. Unlike
classical mechanics, in quantum mechanics it cannot be assumed that the effect
of the measurement on the system can be made arbitrarily small. It is necessary
to supplement quantum theory with additional postulates, describing the
measurement. One of such additional postulate is von Neumann's state reduction
(or projection) postulate \cite{vNeum}. The essential peculiarity of this
postulate is its nonunitary character. However, this postulate refers only to
an ideal measurement, which is instantaneous and arbitrarily accurate. Real
measurements are described by the projection postulate only roughly.

The important consequence of von Neumann's projection postulate is the quantum
Zeno effect. In quantum mechanics short-time behavior of nondecay probability
of unstable particle is not exponential but quadratic \cite{Khalfin}. This
deviation from the exponential decay has been observed by Wilkinson {\it et
al.\/} \cite{GG}. In 1977, Mishra and Sudarshan \cite{Mishra} showed that this
behavior when combined with the quantum theory of measurement, based on the
assumption of the collapse of the wave function, leaded to a very surprising
conclusion: frequent observations slowed down the decay. An unstable particle
would never decay when continuously observed. Mishra and Sudarshan have called
this effect the quantum Zeno paradox or effect. The effect is called so in
allusion to the paradox stated by Greek philosopher Zeno (or Zenon) of Elea.
The very first analysis does not take into account the actual mechanism of the
measurement process involved, but it is based on an alternating sequence of
unitary evolution and a collapse of the wave function. The Zeno effect has been
experimentally proved \cite{Itano} in a repeatedly measured two-level system
undergoing Rabi oscillations. The outcome of this experiment has also been
explained without the collapse hypothesis \cite{Petrosky,FS,Namiki}.

Later it was realized that the repeated measurements could not only slow the
quantum dynamics but the quantum process may be accelerated by frequent
measurements as well
\cite{Kofman1,Kaulakys,Pascazio2,Elattari,Facchi,Kofman2,Lewenstein}. This
effect was called a quantum anti-Zeno effect by Kaulakys and Gontis
\cite{Kaulakys}, who argued that frequent interrogations may destroy quantum
localization effect in chaotic systems. An effect, analogous to the quantum
anti-Zeno effect has been obtained in a computational study involving barrier
penetration, too \cite{Fearn}. Recently, an analysis of the acceleration of a
chemical reaction due to the quantum anti-Zeno effect has been presented in
Ref. \cite{prezhdo}.

Although a great progress in the investigation of the quantum Zeno effect has
been made, this effect is not completely understood as yet. In the analysis of
the quantum Zeno effect the finite duration of the measurement becomes
important, therefore, the projection postulate is not sufficient to solve this
problem. The complete analysis of the Zeno effect requires a more precise model
of measurement than the projection postulate.

The purpose of this article is to consider such a model of the measurement. The
model describes a measurement of the finite duration and finite accuracy.
Although the used model does not describe the irreversible process, it leads,
however, to the correct correlation between the states of the measured system
and the measuring apparatus.

Due to the finite duration of the measurement it is impossible to consider
infinitely frequent measurements, as in Ref. \cite{Mishra}. The highest
frequency of the measurements is achieved when the measurements are performed
one after another, without the period of the measurement-free evolution between
two successive measurements. In this paper we consider such a sequence of the
measurements. Our goal is to check whether this sequence of the measurements
can change the evolution of the system and to verify the predictions of the
quantum Zeno effect.

The work is organized as follows. In section \ref{sec:mod} we present the model
of the measurement. A simple case is considered in section \ref{sec:id} in
order to determine the requirements for the duration of the measurement. In
section \ref{sec:meas} we derived a general formula for the probability of the
jump into another level during the measurement. The effect of repeated
measurements on the system with a discrete spectrum is investigated in section
\ref{sec:discr}. The decaying system is considered in section \ref{sec:dec}.
Section \ref{sec:concl} summarizes our findings.

\section{Model of the measurements}
\label{sec:mod}

We consider a system which consists of two parts. The first part of the system
has the discrete energy spectrum. The Hamiltonian of this part is $\hat{H}_0$.
The other part of the system is represented by Hamiltonian $\hat{H}_1$.
Hamiltonian $\hat{H}_1$ commutes with $\hat{H}_0$. In a particular case the
second part can be absent and $\hat{H}_1$ can be zero. The operator
$\hat{V}(t)$ causes the jumps between different energy levels of $\hat{H}_0$.
Therefore, the full Hamiltonian of the system equals to
$\hat{H}_S=\hat{H}_0+\hat{H}_1+\hat{V}(t)$. The example of such a system is an
atom with the Hamiltonian $\hat{H}_0$ interacting with the electromagnetic
field, represented by $\hat{H}_1$.

We will measure in which eigenstate of the Hamiltonian $\hat{H}_0$ the
system is. The measurement is performed by coupling the system with the
detector. The full Hamiltonian of the system and the detector equals to
\begin{equation}
\hat{H}=\hat{H}_S+\hat{H}_D+\hat{H}_I
\end{equation}
where $\hat{H}_D$ is the Hamiltonian of the detector and $\hat{H}_I$ represents
the interaction between the detector and the system. We choose the operator
$\hat{H}_I$ in the form
\begin{equation}
\hat{H}_I=\lambda\hat{q}\hat{H}_0
\end{equation}
where $\hat{q}$ is the operator acting in the Hilbert space of the detector and
the parameter $\lambda$ describes the strength of the interaction. This
system---detector interaction is that considered by von Neumann \cite{vNeum}
and in Refs. \cite{joos,caves,milb,gagen,rus}. In order to obtain a sensible
measurement, the parameter $\lambda$ must be large. We require a continuous
spectrum of operator $\hat{q}$. For simplicity, we can consider the quantity
$q$ as the coordinate of the detector.

The measurement begins at time moment $t_0$. At the beginning of the
interaction with the detector, the detector is in the pure state $\left|\Phi
\right\rangle$. The full density matrix of the system and detector is
$\hat{\rho}(t_0)=\hat{\rho}_S(t_0)\otimes\left|\Phi\right\rangle\left\langle
\Phi\right|$ where $\hat{\rho}_S(t_0)$ is the density matrix of the system. The
duration of the measurement is $\tau$. After the measurement the density matrix
of the system is $\hat{\rho}_S(\tau+t_0)=Tr_D\left\{\hat{U}(\tau+t_0)\left(
\hat{\rho}_S(t_0)\otimes\left|\Phi\right\rangle\left\langle\Phi\right|\right)
\hat{U}^{\dag}(\tau+t_0)\right\}$ and the density matrix of the detector is
$\hat{\rho}_D(\tau+t_0)=Tr_S\left\{\hat{U}(\tau+t_0)\left(\hat{\rho}_S(t_0)\otimes
\left|\Phi\right\rangle\left\langle\Phi\right|\right)\hat{U}^{\dag}(\tau+t_0)\right\}$
where $\hat{U}(t)$ is the evolution operator of the system and detector,
obeying the equation
\begin{equation}
i\hbar\frac{\partial}{\partial t}\hat{U}(t)=\hat{H}(t)\hat{U}(t)
\end{equation}
with the initial condition $\hat{U}(t_0)=1$.

Since the initial density matrix is chosen in a factorizable form, the density
matrix of the system after the interaction depends linearly on the density
matrix of the system before the interaction. We can represent this fact by the
equality
\begin{equation}
\hat{\rho}_S(\tau+t_0)=S(\tau,t_0)\hat{\rho}_S(t_0)
\label{eq:supop}
\end{equation}
where $S(\tau,t_0)$ is the superoperator acting on the density matrices of the
system. If the vectors $\left|n\right\rangle$ form the complete basis in the
Hilbert space of the system we can rewrite Eq.\ (\ref{eq:supop}) in the form
\begin{equation}
\rho_S(\tau+t_0)_{pr}=S(\tau,t_0)_{pr}^{nm}\rho_S(t_0)_{nm}
\label{eq:rho1}
\end{equation}
where the sum over the repeating indices is supposed. The matrix elements of
the superoperator are
\begin{equation}
S(\tau,t_0)_{pr}^{nm}=Tr_D\left\{\left\langle p\right|\hat{U}
(\tau+t_0)\left(\left|n\right\rangle\left\langle m\right|\otimes
\left|\Phi\right\rangle\left\langle\Phi\right|\right)\hat{U}^{\dag}(\tau+t_0)
\left|r\right\rangle\right\}.
\end{equation}

Due to the finite duration of the measurement it is impossible to realize the
infinitely frequent measurements. The highest frequency of the measurements is
achieved when the measurements are performed one after another without the
period of the measurement-free evolution between two successive measurements.
Therefore, we model a continuous measurement by the subsequent measurements of
the finite duration and finite accuracy. After $N$ measurements the density
matrix of the system is
\begin{equation}
\hat{\rho}_S(N\tau)=S(\tau,(N-1)\tau)\ldots S(\tau,\tau)
S(\tau,0)\hat{\rho}_S(0).
\end{equation}

Further, for simplicity we will neglect the Hamiltonian of the detector. After
this assumption the evolution operator is equal to
$\hat{U}(t,1+\lambda\hat{q})$ where the operator $\hat{U}(t,\xi)$ obeys the
equation
\begin{equation}
i\hbar\frac{\partial}{\partial t}\hat{U}(t,\xi)=\left(\xi
\hat{H}_0+\hat{H}_1+\hat{V}(t+t_0)\right)\hat{U}(t,\xi)
\label{eq:evol1}
\end{equation}
with the initial condition $\hat{U}(t_0,\xi)=1$. Then the superoperator
$S(\tau,t_0)$ is
\begin{equation}
S(\tau,t_0)_{pr}^{nm}=\int dq\left|\langle q\left|\Phi\right\rangle
\right|^2\left\langle p\right|\hat{U}(\tau+t_0,1+\lambda q)\left|n
\right\rangle\left\langle m\right|\hat{U}^{\dag}(\tau+t_0,1+\lambda q)
\left|r\right\rangle.
\label{eq:supmatr}
\end{equation}

\section{Measurement of the unperturbed system}
\label{sec:id}

In order to estimate the necessary duration of the single measurement it is
convenient to consider the case when the operator $\hat{V}=0$. In such a case
the description of the evolution is simpler. The measurement of this kind
occurs also when the influence of the perturbation operator $\hat{V}$ is small
in comparison with the interaction between the system and the detector and,
therefore, the operator $\hat{V}$ can be neglected.

We can choose the basis $\left|n\alpha\right\rangle$ common for the operators
$\hat{H}_0$ and $\hat{H}_1$,
\begin{eqnarray}
\hat{H}_0\left|n\alpha\right\rangle&=&E_n\left|n\alpha\right\rangle\, \\
\hat{H}_1\left|n\alpha\right\rangle&=&E_1(n,\alpha)\left|n\alpha\right\rangle
\end{eqnarray}
where $n$ numbers the eigenvalues of the Hamiltonian $\hat{H}_0$ and $\alpha$
represents the remaining quantum numbers. Since the Hamiltonian of the system
does not depend on $t$ we will omit the parameter $t_0$ in this section. From
Eq.\ (\ref{eq:supmatr}) we obtain the superoperator $S(\tau)$ in the basis
$\left|n\alpha\right\rangle$
\begin{eqnarray}
S(\tau)_{p\alpha_3,r\alpha_4}^{n\alpha_1,m\alpha_2}&=&\delta_{np}
\delta_{mr}\delta(\alpha_1,\alpha_3)\delta(\alpha_2,\alpha_4)
\exp(i\omega_{m\alpha_2,n\alpha_1}\tau)\nonumber \\
&&\times\int dq\left|\langle q\left|\Phi\right\rangle\right|^2\exp(
i\lambda\omega_{mn}\tau q)
\label{eq:s1}
\end{eqnarray}
where
\begin{eqnarray}
\omega_{mn}&=&\frac{1}{\hbar}(E_m-E_n)\,, \\
\omega_{m\alpha_2,n\alpha_1}&=&\omega_{mn}+\frac{E_1(m,\alpha_2)-E_1(
n,\alpha_1)}{\hbar}
\end{eqnarray}
and $\delta({\cdot},{\cdot})$ represent the Kronecker's delta in a discrete
case and the Dirac's delta in a continuous case. Eq.\ (\ref{eq:s1}) can be
rewritten using the correlation function
\begin{equation}
F(\nu)=\left\langle\Phi\right|\exp(i\nu\hat{q})\left|\Phi\right\rangle.
\label{F}
\end{equation}
We can express this function as $F(\nu)=\int dq\left|\langle q\left|\Phi
\right\rangle\right|^2\exp(i\nu q)=\int dp\left\langle\Phi\right|p\rangle
\langle p-\frac{\nu}{\hbar}\left|\Phi\right\rangle$. Since vector $\left|
\Phi\right\rangle$ is normalized, the function $F(\nu)$ tends to zero when
$\left|\nu\right|$ increases. There exists a constant $C$ such that the
correlation function $\left|F(\nu)\right|$ is small if the variable
$\left|\nu\right|>C$.

Then the equation for the superoperator $S(\tau)$ is
\begin{equation}
S(\tau)_{p\alpha_3,r\alpha_4}^{n\alpha_1,m\alpha_2}=\delta_{np}
\delta_{mr}\delta(\alpha_1,\alpha_3)\delta(\alpha_2,\alpha_4)
\exp(i\omega_{m\alpha_2,n\alpha_1}\tau)F(\lambda \tau\omega_{mn}).
\label{eq:s2}
\end{equation}
Using Eqs.\ (\ref{eq:rho1}) and\ (\ref{eq:s2}) we find that after the
measurement the non-diagonal elements of the density matrix of the system
become small, since $F(\lambda \tau\omega_{mn})$ is small for $n\neq m$ when
$\lambda \tau$ is large.

The density matrix of the detector is
\begin{equation}
\left\langle q\right|\hat{\rho}_D(\tau)\left|q_1\right\rangle=\langle
q\left|\Phi\right\rangle\left\langle\Phi\right|q_1\rangle Tr\left\{
\hat{U}(\tau,1+\lambda q)\hat{\rho}_S(0)\hat{U}^{\dag}(\tau,1+\lambda
q_{1})\right\}.\label{eq:det1}
\end{equation}
From Eqs.\ (\ref{eq:evol1}) and\ (\ref{eq:det1}) we obtain
\begin{equation}
\left\langle q\right|\hat{\rho}_D(\tau)\left|q_1\right\rangle=\langle
q\left|\Phi \right\rangle\left\langle\Phi\right|q_1\rangle\sum_n\exp(i\lambda
\tau\omega_n(q_1-q))\sum_{\alpha}\left\langle n,\alpha\right|
\hat{\rho}_S(0)\left|n,\alpha\right\rangle
\end{equation}
where
\begin{equation}
\omega_n=\frac{1}{\hbar}E_n.
\end{equation}
The probability that the system is in the energy level $n$ may be expressed as
\begin{equation}
P(n)=\sum_{\alpha}\left\langle n,\alpha\right|\hat{\rho}_S(0)\left|n,\alpha
\right\rangle.
\end{equation}
Introducing the state vectors of the detector
\begin{equation}
\left|\Phi_E\right\rangle=\exp\left(-\frac{i}{\hbar}\lambda \tau
E\hat{q}\right) \left|\Phi\right\rangle
\end{equation}
we can express the density operator of the detector as
\begin{equation}
\hat{\rho}_D(\tau)=\sum_n\left|\Phi_{E_n}\right\rangle\left\langle\Phi_{E_n}\right|
P(n).
\end{equation}

The measurement is complete when the states $\left|\Phi_E\right\rangle$ are
almost orthogonal. The different energies can be separated only when the
overlap between the corresponding states $\left|\Phi_E\right\rangle$ is almost
zero. The scalar product of the states $\left|\Phi_E\right\rangle$ with
different energies $E_1$ and $E_2$ is
\begin{equation}
\langle\Phi_{E_1}\left|\Phi_{E_2}\right\rangle=F(\lambda \tau\omega_{12}).
\end{equation}
The correlation function $\left|F(\nu)\right|$ is small when
$\left|\nu\right|>C$. Therefore, we have the estimation for the error of the
energy measurement $\Delta E$ as
\begin{equation}
\lambda \tau\Delta E\gtrsim\hbar C
\end{equation}
and we obtain the expression for the necessary duration of the measurement
\begin{equation}
\tau\gtrsim\frac{\hbar}{\Lambda\Delta E} \label{eq:tt}
\end{equation}
where
\begin{equation}
\Lambda=\frac{\lambda}{C}\,.
\end{equation}
Since in our model the measurements are performed immediately one after the
other, from Eq.~(\ref{eq:tt}) it follows that the rate of measurements is
proportional to the strength of the interaction $\lambda$ between the system
and the measuring device.

\section{Measurement of the perturbed system}
\label{sec:meas}

The operator $\hat{V}(t)$ represents the perturbation of the unperturbed
Hamiltonian $\hat{H}_0+\hat{H}_1$. We will take into account the influence of
the operator $\hat{V}$ by the perturbation method, assuming that the strength
of the interaction $\lambda$ between the system and detector is large.

The operator $\hat{V}(t)$ in the interaction picture is
\begin{equation}
\tilde{V}(t,t_0,\xi)=\exp\left(\frac{i}{\hbar}(\xi\hat{H}_0+\hat{H}_1)t\right)
\hat{V}(t+t_0)\exp\left(-\frac{i}{\hbar}(\xi\hat{H}_0+\hat{H}_1)t\right).
\end{equation}
In the second order approximation the evolution operator equals to
\begin{eqnarray}
\hat{U}(\tau,t_0,\xi)&\approx&\exp\left(-\frac{i}{\hbar}(\xi\hat{H}_0+\hat{H}_1)\tau
\right)\left\{1+\frac{1}{i\hbar}\int_0^\tau dt\tilde{V}
(t,t_0,\xi)\right.\nonumber \\
&&-\left.\frac{1}{\hbar^2}\int_0^\tau dt_1\int_0^t
dt_2\tilde{V}(t_1,t_0,\xi)\tilde{V}(t_2,t_0\xi)\right\}.
\label{eq:evol2}
\end{eqnarray}
Using Eqs.\ (\ref{eq:supmatr}) and\ (\ref{eq:evol2}) we can obtain the
superoperator $S$ in the second order approximation, too. The expression for
the matrix elements of the superoperator $S$ is given in the appendix (Eqs.\
(\ref{eq:ap1}),\ (\ref{eq:ap2}) and\ (\ref{eq:ap3})).

The probability of the jump from the level $|i\alpha\rangle$ to the level
$|f\alpha_1\rangle$ during the measurement is $W(i\alpha\rightarrow f\alpha_1)
=S(\tau,t_0)_{f\alpha_1,f\alpha_1}^{i\alpha,i\alpha}$. Using Eqs.\
(\ref{eq:ap1}),\ (\ref{eq:ap2}) and\ (\ref{eq:ap3}) we obtain
\begin{eqnarray}
W(i\alpha\rightarrow f\alpha_1)&=&\frac{1}{\hbar^2}\int_0^\tau dt_1\int_0^\tau
dt_2 F(\lambda\omega_{if}(t_2-t_1))V(t_1+t_0)_{f\alpha_1,i\alpha}
V(t_2+t_0)_{i\alpha,f\alpha_1}\nonumber \\
&&\times\exp\left(i\omega_{i\alpha,f\alpha_1}(t_2-t_1)\right).
\label{eq:W1}
\end{eqnarray}

The expression for the jump probability can be further simplified if the
operator $\hat{V}$ does not depend on $t$. We introduce the function
\begin{equation}
\Phi(t)_{f\alpha_1,i\alpha}=\left|V_{f\alpha_1,i\alpha}\right|^2
\exp\left(\frac{i}{\hbar}(E_1(f,\alpha_1)-E_1(i,\alpha))t\right).
\end{equation}
Changing variables we can rewrite the jump probability as
\begin{equation}
W(i\alpha\rightarrow f\alpha_1)=\frac{2}{\hbar^2}\text{Re}\int_0^\tau dt
F(\lambda\omega_{fi}t)\exp(i\omega_{fi}t)(\tau-t) \Phi(t)_{f\alpha_1,i\alpha}.
\label{eq:W}
\end{equation}
Introducing the Fourier transformation of $\Phi(t)_{f\alpha_1,i\alpha}$
\begin{equation}
G(\omega)_{f\alpha_1,i\alpha}=\frac{1}{2\pi}\int_{-\infty}^{\infty}dt
\Phi(t)_{f\alpha_1,i\alpha}\exp(-i\omega t)
\end{equation}
and using Eq.\ (\ref{eq:W}) we obtain the equality
\begin{equation}
W(i\alpha\rightarrow f\alpha_1)=\frac{2\pi\tau}{\hbar^2}\int_{-\infty}^{\infty}
d\omega G(\omega)_{f\alpha_1,i\alpha}P(\omega)_{if}
\label{eq:resW}
\end{equation}
where
\begin{equation}
P(\omega)_{if}=\frac{1}{\pi}\text{Re}\int_0^\tau dt
F(\lambda\omega_{if}t)\exp(i(\omega-\omega_{if})t)
\left(1-\frac{t}{\tau}\right). \label{eq:pp}
\end{equation}
From Eq.\ (\ref{eq:pp}), using the equality $F(0)=1$, we obtain
\begin{equation}
\int d\omega P(\omega)_{if}=1.
\label{eq:prop3}
\end{equation}

The quantity $G$ equals to
\begin{equation}
G(\omega)_{f\alpha_1,i\alpha}=\hbar\left|V_{f\alpha_1,i\alpha}\right|^2
\delta(E_1(f,\alpha_1)-E_1(i,\alpha)-\hbar\omega).
\label{eq:G}
\end{equation}
We see that the quantity $G(\omega)$ characterizes the perturbation.

\section{The discrete spectrum}
\label{sec:discr}

Let us consider the measurement effect on the system with the discrete
spectrum. The Hamiltonian $\hat{H}_0$ of the system has a discrete spectrum,
the operator $\hat{H}_1=0$, and the operator $\hat{V}(t)$ represents a
perturbation resulting in the quantum jumps between the discrete states of the
system $\hat{H}_0$.

For the separation of the energy levels, the error in the measurement should be
smaller than the distance between the nearest energy levels of the system. It
follows from this requirement and Eq.\ (\ref{eq:tt}) that the measurement time
$\tau\gtrsim\frac{1}{\Lambda\omega_{\min}}$, where $\omega_{\min}$ is the
smallest of the transition frequencies $|\omega_{if}|$.

When $\lambda$ is large then $\left|F(\lambda x)\right|$ is not very small only
in the region $\left|x\right|<\Lambda^{-1}$. We can estimate the probability of
the jump to the other energy level during the measurement, replacing $F(\nu)$
by $2C\delta(\nu)$in Eq.\ (\ref{eq:W1}). Then from Eq.\ (\ref{eq:W1}) we obtain
\begin{equation}
W(i\alpha\rightarrow f\alpha_1)\approx
\frac{2}{\hbar^2\Lambda\left|\omega_{if}\right|}\int_0^\tau dt \left|
V(t+t_0)_{i\alpha_1,f\alpha}\right|^2. \label{eq:jumpprob}
\end{equation}
We see that the probability of the jump is proportional to $\Lambda^{-1}$.
Consequently, for large $\Lambda$, i.e. for the strong interaction with the
detector, the jump probability is small. This fact represents the quantum Zeno
effect. However, due to the finiteness of the interaction strength the jump
probability is not zero. After sufficiently large number of measurements the
jump occurs. We can estimate the number of measurements $N$ after which the
system jumps into other energy levels from the equality
$\frac{2\tau}{\hbar^2\Lambda\left|\omega_{\min}\right|}\left|V_{\max}
\right|^2N\sim 1$ where $|V_{\max}|$ is the largest matrix element of the
perturbation operator $V$. This estimation allows us to introduce the
characteristic time, during which the evolution of the system is inhibited
\begin{equation}
t_{\text{inh}}\equiv \tau
N=\Lambda\frac{\hbar^2\left|\omega_{\min}\right|}{2\left| V_{\max}\right|^2}\,.
\label{eq:inht}
\end{equation}
We call this duration the inhibition time (it is natural to call this duration
the Zeno time, but this term has already different meaning).

The full probability of the jump from level $|i\alpha\rangle$ to other levels
is $W(i\alpha)=\sum_{f,\alpha_1}W(i\alpha\rightarrow f\alpha_1)$. From Eq.\
(\ref{eq:jumpprob}) we obtain
\begin{equation}
W(i\alpha)=\frac{2}{\hbar^2\Lambda}\sum_{f,\alpha_1}
\frac{1}{\left|\omega_{if}\right|}\int_0^\tau dt
\left|V\left(t+t_0\right)_{f\alpha_1,i\alpha}\right|^2.
\end{equation}
If the matrix elements of the perturbation $V$ between different levels are of
the same size then the jump probability increases linearly with the number of
the energy levels. This behavior has been observed in Ref. \cite{Gagen2}.

Due to the unitarity of the operator $\hat{U}(t,\xi)$ it follows from Eq.\
(\ref{eq:supmatr}) that the superoperator $S(\tau,t_0)$ obeys the equalities
\begin{mathletters}
\begin{eqnarray}
\sum_{p,\alpha}S(\tau,t_0)_{p\alpha,p\alpha}^{n\alpha_1,m\alpha_2}
&=&\delta_{nm}\delta_{\alpha_1,\alpha_2}, \\
\sum_{n,\alpha}S(\tau,t_0)_{p\alpha_1,r\alpha_2}^{n\alpha,n\alpha}
&=&\delta_{pr}\delta_{\alpha_1,\alpha_2}. \label{eq:prop1}
\end{eqnarray}
\end{mathletters}
If the system has a finite number of energy levels, the density matrix of the
system is diagonal and all states are equally occupied (i.e.,
$\rho(t_0)_{n\alpha_1,m\alpha_2}=\frac{1}{K}
\delta_{nm}\delta_{\alpha_1,\alpha_2}$ where $K$ is the number of the energy
levels) then from Eq.\ (\ref{eq:prop1}) it follows that
$S(\tau,t_0)\rho(t_0)=\rho(t_0)$. Such a density matrix is the stable point of
the map $\rho\rightarrow S\rho$. Therefore, we can expect that after a large
number of measurements the density matrix of the system tends to this density
matrix.

When $\Lambda$ is large and the duration of the measurement is small, we can
neglect the non-diagonal elements in the density matrix of the system, since
they always are of order $\Lambda^{-1}$. Replacing $F(\nu)$ by $2C\delta(\nu)$
in Eqs.\ (\ref{eq:ap1}),\ (\ref{eq:ap2}) and\ (\ref{eq:ap3}) and neglecting the
elements of the superoperator $S$ that cause the arising of the non-diagonal
elements of the density matrix, we can write the equation for the
superoperator $S$ as
\begin{equation}
S(\tau,t_0)_{p\alpha_3,r\alpha_4}^{n\alpha_1,m\alpha_2}\approx
\delta_{pn}\delta(\alpha_3,\alpha_1)\delta_{rm}
\delta(\alpha_4,\alpha_2)\delta_{pr}+\frac{1}{\Lambda}
A(\tau,t_0)_{p,\alpha_3,\alpha_4}^{n,\alpha_1,\alpha_2}\delta_{pr}\delta_{nm}
\end{equation}
where
\begin{eqnarray}
A(\tau,t_0)_{p,\alpha_3,\alpha_4}^{n,\alpha_1,\alpha_2}
&=&\frac{2}{\hbar^2\left|\omega_{np}\right|}\int_0^\tau dt
V(t+t_0)_{p\alpha_3,n\alpha_1}V(t+t_0)_{n\alpha_2,p\alpha_4}\nonumber \\
&&-\delta_{pn}\delta(\alpha_4,\alpha_2)\sum_{s,\alpha}
\frac{1}{\hbar^2\left|\omega_{sn}\right|}\int_0^\tau dt
V(t+t_0)_{n\alpha_3,s\alpha}V(t+t_0)_{s\alpha,n\alpha_1}
\nonumber \\
&&-\delta_{pn}\delta(\alpha_3,\alpha_1)\sum_{s,\alpha}
\frac{1}{\hbar^2\left|\omega_{ns}\right|}\int_0^\tau dt V(t+t_0
)_{s\alpha,n\alpha_4}V(t+t_0)_{n\alpha_2,s\alpha}
\end{eqnarray}
Then for the diagonal elements of the density matrix we have
$\rho(\tau+t_0)\approx\rho(t_0)+\frac{1}{\Lambda}A(\tau,t_0)\rho(t_0)$, or
\begin{equation}
\frac{d}{dt}\hat{\rho}(t)\approx\frac{1}{\Lambda\tau}A(\tau,t) \hat{\rho}(t).
\label{eq:deriv}
\end{equation}
If the perturbation $V$ does not depend on $t$ then it follows from
Eq.~(\ref{eq:deriv}) that the diagonal elements of the density matrix evolve
exponentially.

\subsection{Example}

As an example we will consider the evolution of the measured two-level system.
The system is forced by the perturbation $V$ which induces the jumps from one
state to another. The Hamiltonian of this system is
\begin{equation}
\hat{H}=\hat{H}_0+\hat{V} \label{eq:ham1}
\end{equation}
where
\begin{eqnarray}
\hat{H}_0&=&\frac{\hbar\omega}{2}\hat{\sigma}_3, \label{eq:ham2}\\
\hat{V}&=&v\hat{\sigma}_{+}+v^{*}\hat{\sigma}_{-}. \label{eq:ham3}
\end{eqnarray}
Here $\sigma_1,\sigma_2,\sigma_3$ are Pauli matrices and $\sigma_{\pm}
=\frac{1}{2}(\sigma_1\pm i\sigma_2)$. The Hamiltonian $\hat{H}_0$ has two
eigenfunctions $\left|0\right\rangle$ and $\left|1\right\rangle$ with the
eigenvalues $-\hbar\frac{\omega}{2}$ and $\hbar\frac{\omega}{2}$ respectively.
The evolution operator of the unmeasured system is
\begin{equation}
\hat{U}(t)=\cos\left(\frac{\Omega}{2}t\right)-\frac{2i}{\hbar\Omega}\hat{H}
\sin\left(\frac{\Omega}{2}t\right)
\label{eq:evol4}
\end{equation}
where
\begin{equation}
\Omega=\sqrt{\omega^2+4\frac{\left|v\right|^2}{\hbar^2}}.
\end{equation}
If the initial density matrix is $\rho(0)=|1\rangle\langle 1|$ then the
evolution of the diagonal elements of the unmeasured system's density matrix is
given by the equations
\begin{mathletters}
\begin{eqnarray}
\rho_{11}(t)&=&\cos^2\left(\frac{\Omega}{2}
t\right)+\left(\frac{\omega}{\Omega}\right)^2\sin^2\left(\frac{\Omega}{2}t\right)
\label{eq:free}\\
\rho_{00}(t)&=&\left(1-\left(\frac{\omega}{\Omega}\right)^2\right)\sin^2\left(
\frac{\Omega}{2}t\right).
\end{eqnarray}
\end{mathletters}

Let us consider now the dynamics of the measured system. The equations for the
diagonal elements of the density matrix (Eq.\ (\ref{eq:deriv}) ) for the system
under consideration are
\begin{mathletters}
\label{eq:eqs1}
\begin{eqnarray}
\frac{d}{dt}\rho_{11} &\approx &-\frac{1}{t_{\text{inh}}}\left(\rho_{11}
-\rho_{00}\right), \\
\frac{d}{dt}\rho_{00} &\approx &-\frac{1}{t_{\text{inh}}}\left(\rho_{00}
-\rho_{11}\right).
\end{eqnarray}
\end{mathletters}
where the inhibition time, according to Eq.\ (\ref{eq:inht}), is
\begin{equation}
t_{\text{inh}}=\frac{\Lambda}{2\omega}\left|\frac{\hbar\omega}{v}\right|^2.
\end{equation}
The solution of Eqs.\ (\ref{eq:eqs1}) with the initial condition
$\rho(0)=\left|1\right\rangle\left\langle 1\right|$ is
\begin{mathletters}
\begin{eqnarray}
\rho_{11}(t) &=&\frac{1}{2}\left(1+\exp\left(-\frac{2}{t_{
\text{inh}}}t\right)\right), \label{eq:aprox1}\\
\rho_{00}(t) &=&\frac{1}{2}\left(1-\exp\left(-\frac{2}{t_{
\text{inh}}}t\right)\right).
\end{eqnarray}
\end{mathletters}

From Eq.~(\ref{eq:prop1}) it follows that if the density matrix of the system
is
\begin{equation}
\hat{\rho}_f=\frac{1}{2}\left(\left|0\right\rangle\left\langle 0\right|
+\left|1\right\rangle\left\langle 1\right|\right),
\end{equation}
then $S(\tau)\hat{\rho}_f=\hat{\rho}_f$. Hence, when the number of the
measurements tends to infinity, the density matrix of the system approaches
$\hat{\rho}_f$.

\begin{figure}
  \begin{center}
    \epsfxsize=.6\hsize
    \epsffile{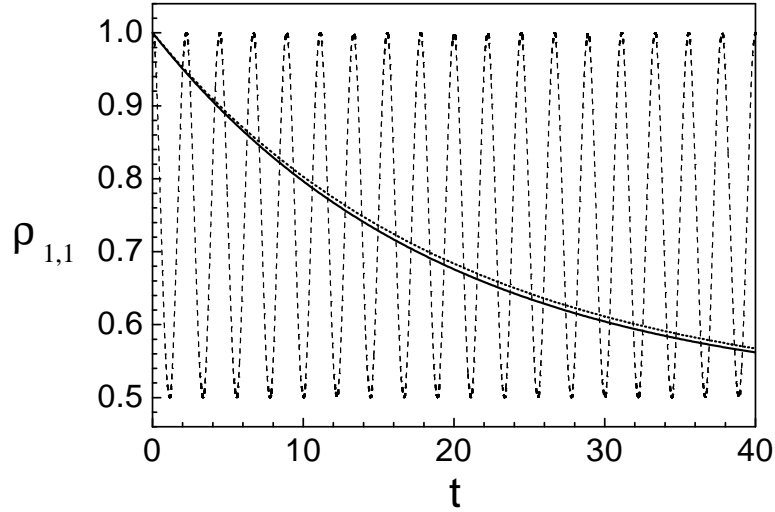}
  \end{center}
\caption{The occupation of the initial level $1$ of the measured two-level
system calculated according to Eqs.\ (\protect\ref{eq:rho1}),\
(\protect\ref{eq:supmatr}),\ (\protect\ref{eq:evol4}) and\
(\protect\ref{eq:fcalc}). The used parameters are $\hbar=1$, $\sigma^2=1$,
$\omega=2$, $v=1$. The strength of the measurement $\lambda=50$ and the
duration of the measurement $\tau=0.1$. The exponential approximation
(\protect\ref{eq:aprox1}) is shown as a dashed line. For comparison the
occupation of the level $1$ of the unmeasured system is also shown (dotted
line).}
\label{fig1}
\end{figure}
\begin{figure}
  \begin{center}
    \epsfxsize=.6\hsize
    \epsffile{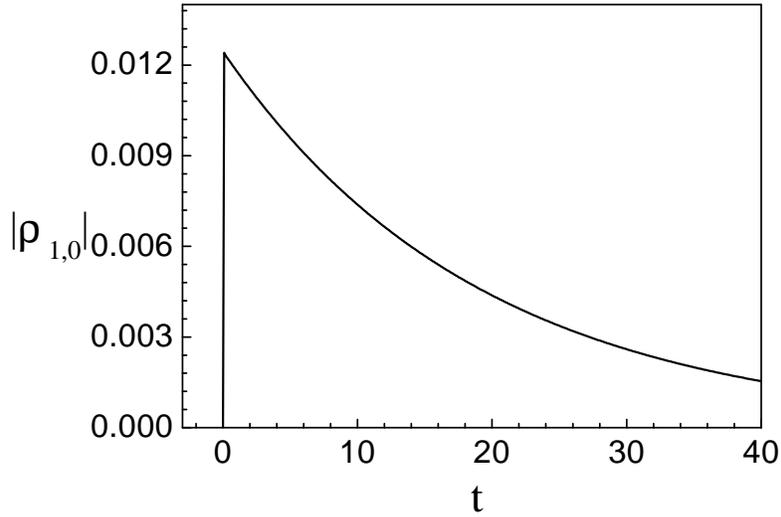}
  \end{center}
\caption{The non-diagonal element of the density matrix of the measured
two-level system. Used parameters are the same as in Fig. \protect\ref{fig1}}
\label{fig2}
\end{figure}
\begin{figure}
  \begin{center}
    \epsfxsize=.6\hsize
    \epsffile{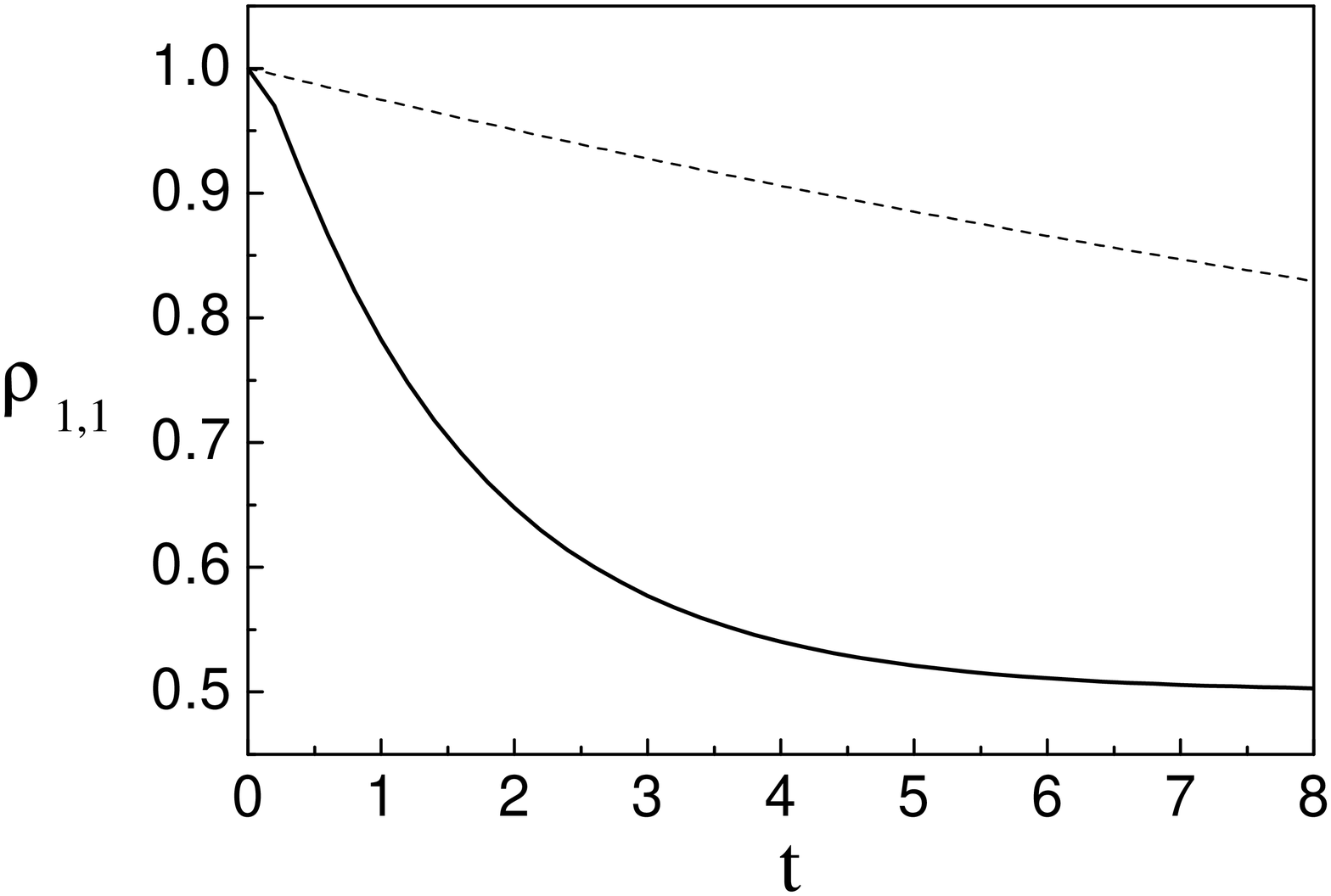}
  \end{center}
\caption{The occupation of the initial level $1$ of the measured two-level
system for different strengths of the measurement: $\lambda=50$, $\tau=0.1$
(dashed line) and $\lambda=5$, $\tau=0.2$ (solid line). Other parameters are
the same as in Fig. \protect\ref{fig1}}
\label{fig3}
\end{figure}

We have performed the numerical analysis of the dynamics of the measured
two-level system (\ref{eq:ham1})---(\ref{eq:ham3}) using Eqs.\
(\ref{eq:rho1}),\ (\ref{eq:supmatr}) and\ (\ref{eq:evol4}) with the Gaussian
correlation function (\ref{F})
\begin{equation}
F(\nu)=\exp\left(-\frac{\nu^2}{2\sigma^2}\right). \label{eq:fcalc}
\end{equation}
From the condition $\int_{-\infty}^{\infty}F(\nu)d\nu=2C$ we have
$C=\sigma\sqrt{\frac{\pi}{2}}$. The initial state of the system is
$\left|1\right\rangle$. The matrix elements of the density matrix
$\rho(t)_{11}$ and $\rho(t)_{10}$ are represented in Fig. \ref{fig1} and Fig.
\ref{fig2}, respectively. In Fig. \ref{fig1} the approximation
(\ref{eq:aprox1}) is also shown. This approach is close to the exact evolution.
The matrix element $\rho(t)_{11}$ for two different values of $\lambda$ is
shown in Fig. \ref{fig3}. We see that for larger $\lambda$ the evolution of the
system is slower.

The influence of the repeated non-ideal measurements on the two level system
driven by the periodic perturbation has also been considered in Refs.
\cite{Gagen2,Peres,Jordan,Venugoplan}. Similar results have been found: the
occupation of the energy levels changes exponentially with time, approaching
the limit $\frac{1}{2}$.

\section{The decaying system}
\label{sec:dec}

We consider the system which consists of two parts. We can treat the first part
as an atom, and the second part as the field (reservoir). The energy spectrum
of the atom is discrete and the spectrum of the field is continuous. The
Hamiltonians of these parts are $\hat{H}_0$ and $\hat{H}_1$ respectively and
the eigenfunctions are $\left|n\right\rangle$ and $\left|\alpha\right\rangle$,
\begin{mathletters}
\begin{eqnarray}
\hat{H}_0\left|n\right\rangle&=&E_{n}\left|n\right\rangle,  \\
\hat{H}_1\left|\alpha\right\rangle&=&E_{\alpha}\left|\alpha\right\rangle.
\end{eqnarray}
\end{mathletters}
There is the interaction between the atom and the field represented by the
operator $\hat{V}$. So, the Hamiltonian of the system is
\begin{equation}
\hat{H}_{S}=\hat{H}_{0}+\hat{H}_{1}+\hat{V}.
\end{equation}
The basis for the full system is $\left|n\alpha\right\rangle=\left|
n\right\rangle\otimes\left|\alpha\right\rangle$.

When the measurement is not performed, such a system exhibits exponential
decay, valid for the intermediate times. The decay rate is given according to
the Fermi's Golden Rule
\begin{equation}
R(i\alpha_1\rightarrow f\alpha_2)=\frac{2\pi}{\hbar}
\left|V_{f\alpha_2,i\alpha_1}\right|^2\rho\left(\hbar\omega_{if}\right)
\label{eq:goldrule}
\end{equation}
where
\begin{equation}
\frac{1}{\hbar}(E_{\alpha_2}-E_{\alpha_1})=\omega_{if}
\end{equation}
and $\rho\left(E\right)$ is the density of the reservoir's states.

When the energy level of the atom is measured, we can use the perturbation
theory, as it is in the discrete case.

The initial state of the field is a vacuum state $\left|0\right\rangle$
with energy $E_0=0$. Then the density matrix of the atom is $\hat{\rho}_0
(\tau)=Tr_1\left\{\hat{\rho}(\tau)\right\}=Tr_1
\left\{S(\tau)\hat{\rho}(0)\right\}$ or
$\hat{\rho}_0(\tau)=S_{ef}(\tau)\hat{\rho}_0(0)$, where $S_{ef}$ is an
effective superoperator
\begin{equation}
S_{ef}(\tau)_{pr}^{nm}=\sum_{\alpha}S(\tau)_{p\alpha,r\alpha}^{n0,m0}.
\label{eq:sef}
\end{equation}
When the states of the atom are weakly coupled to a broad band of states
(continuum), the transitions back to the excited state of the atom can be
neglected (i.e., we neglect the influence of emitted photons on the atom).
Therefore, we can use the superoperator $S_{ef}$ for determination of the
evolution of the atom.

Since the states in the reservoir are very dense, one can replace the sum over
$\alpha $ by an integral over $E_{\alpha}$
\[
\sum_{\alpha}\ldots=\int dE_{\alpha}\rho(E_{\alpha})\ldots
\]
where $\rho(E_{\alpha})$ is the density of the states in the reservoir.

\subsection{The spectrum}

The density matrix of the field is $\hat{\rho}_1(\tau)=Tr_0\left\{
\hat{\rho}(\tau)\right\}=Tr_0\left\{S(\tau)\hat{\rho}(0)\right\}$. The diagonal
elements of the field's density matrix give the spectrum. If the initial state
of the atom is $\left|i\right\rangle$ then the distribution of the field's
energy is $W\left(E_{\alpha}\right)=\rho_1(\tau)_{\alpha\alpha}=
\sum_{f}S(\tau)_{f\alpha,f\alpha}^{i0,i0}$ . From Eqs.\ (\ref{eq:ap1}),\
(\ref{eq:ap2}) and\ (\ref{eq:ap3}) we obtain
\begin{equation}
W(E_{\alpha})=\sum_{f}\frac{2\pi}{\hbar^2}\left|V_{f\alpha,i0}\right|^2\tau
P\left(\frac{E_{\alpha}}{\hbar}\right)_{if}
\label{eq:spectr}
\end{equation}
where $P(\omega)_{if}$ is given by the equation (\ref{eq:pp}).
From Eq.\ (\ref{eq:spectr}) we see that $P(\omega)$ is the measurement-modified
shape of the spectral line.

The integral in Eq.\ (\ref{eq:pp}) is small when the exponent oscillates more
rapidly than the function $F$. This condition is fulfilled when
$\frac{E}{\hbar}-\omega_{if}\gtrsim\frac{\lambda\omega_{if}}{C}$. Consequently,
the width of the spectral line is
\begin{equation}
\Delta E_{if}=\Lambda\hbar\omega_{if}.
\end{equation}
The width of the spectral line is proportional to the strength of the
measurement (this equation is obtained using the assumption that the strength
of the interaction with the measuring device $\lambda$ is large and, therefore,
the natural width of the spectral line can be neglected). The broadening of the
spectrum of the measured system is also reported in Ref. \cite{Elattari} for
the case of an electron tunneling out of a quantum dot.

\subsection{The decay rate}

The probability of the jump from the state $i$ to the state $f$ is
$W(i\rightarrow f;\tau)=S_{ef}(\tau)_{ff}^{ii}$. From Eqs.\ (\ref{eq:sef}) it
follows
\begin{equation}
W(i\rightarrow f;\tau)=\sum_{\alpha}W(i0,\rightarrow f\alpha,\tau)
\end{equation}
Using Eq.\ (\ref{eq:resW}) we obtain the equality
\begin{equation}
W\left(i\rightarrow f;\tau\right)=\frac{2\pi
\tau}{\hbar^{2}}\int_{-\infty}^{+\infty} d\omega
G\left(\omega\right)_{fi}P\left(\omega\right)_{if}.
\label{eq:decprob2}
\end{equation}
where
\begin{equation}
G(\omega)_{fi}=\int dE_{\alpha}\rho(E_{\alpha})G(\omega)_{f\alpha,i0}
\end{equation}
The expression for $G(\omega)$ according to Eq.\ (\ref{eq:G}) is
\begin{equation}
G(\omega)_{fi}=\hbar\rho(\hbar\omega)\left|V_{fE_{\alpha}=\hbar\omega
,i0}\right|^2.
\end{equation}
The quantity $G\left(\omega\right)$ is the reservoir coupling spectrum.

The measurement-modified decay rate is $R\left(i\rightarrow f\right)=
\frac{1}{\tau}W(i\rightarrow f;\tau)$. From Eq.\ (\ref{eq:decprob2}) we have
\begin{equation}
R(i\rightarrow f)=\frac{2\pi}{\hbar^2}\int_{-\infty}^{\infty}d\omega
G(\omega)_{fi}P(\omega)_{if}. \label{eq:result}
\end{equation}
The equation\ (\ref{eq:result}) represents a universal result: the decay rate
of the frequently measured decaying system is determined by the overlap of the
reservoir coupling spectrum and the measurement-modified level width. This
equation was derived by Kofman and Kurizki \cite{Kofman2}, assuming the ideal
instantaneous projections. We show that Eq.\ (\ref{eq:result}) is valid for the
more realistic model of the measurement, as well. An equation, similar to Eq.\
(\ref{eq:result}) has been obtained in Ref. \cite{Panov}, considering a
destruction of the final decay state.

Depending on the reservoir spectrum $G(\omega)$ and the strength of the
measurement the inhibition or acceleration of the decay can be obtained. If the
interaction with the measuring device is weak and, consequently, the width of
the spectral line is much smaller than the width of the reservoir spectrum, the
decay rate equals the decay rate of the unmeasured
system, given by the Fermi's Golden Rule\ (\ref{eq:goldrule}). In the
intermediate region, when the width of the spectral line is rather small
compared with the distance between $\omega_{if}$ and the nearest maximum in
the reservoir spectrum, the decay rate grows with increase of $\Lambda$. This
results in the anti-Zeno effect.

If the width of the spectral line is much greater compared both with the width
of the reservoir spectrum and the distance between $\omega_{if}$ and the
centrum of the reservoir spectrum, the decay rate decreases when $\Lambda$
increases. This results in the quantum Zeno effect. In such a case we can use
the approximation
\begin{equation}
G\left(\omega\right)_{fi}\approx\hbar B_{fi}\delta\left(\omega-\omega_R
\right).
\end{equation}
where $B_{fi}$ is defined by the equality $B_{fi}=\frac{1}{\hbar}\int
G\left(\omega\right)_{fi}d\omega$ and $\omega_R$ is the centrum of $G(\omega)$.
Then from Eq.\ (\ref{eq:result}) we obtain the decay rate $R\left(i\rightarrow
f\right)\approx\frac{2\pi}{\hbar}B_{fi} P\left(\omega_{if}\right)_{if}$. From
Eq.\ (\ref{eq:pp}), using the condition $\Lambda\tau
\left|\omega_{if}\right|\gg 1$ and the equality
$\int_{-\infty}^{\infty}F\left(\nu\right)d\nu=2C$ we obtain
\begin{equation}
P\left(\omega_{if}\right)_{if}=\frac{1}{\pi\Lambda\omega_{if}}.
\end{equation}
Therefore, the decay rate is equal to
\begin{equation}
R\left(i\rightarrow f\right)\approx\frac{2 B_{fi}}{\Lambda\hbar\omega_{if}}.
\end{equation}
The obtained decay rate is insensitive to the spectral shape of the reservoir
and is inverse proportional to the measurement strength $\Lambda$.

\section{Summary}
\label{sec:concl}

In this work we investigate the quantum Zeno effect using the definite model of
the measurement. We take into account the finite duration and the finite
accuracy of the measurement. The general equation for the probability of the
jump during the measurement is derived (\ref{eq:resW}). The behavior of the
system under the repeated measurements depends on the strength of measurement
and on the properties of the system.

When the the strength of the interaction with the measuring device is
sufficiently large, the frequent measurements of the system with discrete
spectrum slow down the evolution. However, the evolution cannot be fully
stopped. Under the repeated measurements the occupation of the energy levels
changes exponentially with time, approaching the limit of the equal occupation
of the levels. The jump probability is inversely proportional to the strength
of the interaction with the measuring device.

In the case of a continuous spectrum the measurements can cause inhibition or
acceleration of the evolution. Our model of the continuous measurement gives
the same result as the approach based on the projection postulate
\cite{Kofman2}. The decay rate is equal to the convolution of the reservoir
coupling spectrum with the measurement-modified shape of the spectral line. The
width of the spectral line is proportional to the strength of the interaction
with the measuring device. When this width is much greater than the width of
the reservoir, the quantum Zeno effect takes place. Under these conditions the
decay rate is inversely proportional to the strength of the interaction with
the measuring device. In a number of decaying systems, however, the reservoir
spectrum $G(\omega)$ grows with frequency almost up to the relativistic cut-off
and the strength of the interaction required for the appearance of the quantum
Zeno effect is so high that the initial system is significantly modified. When
the spectral line is not very broad, the decay rate may be increased by the
measurements more often than it may be decreased and the quantum anti-Zeno
effect can be obtained.

\appendix
\section*{The superoperator}

We obtain the superoperator $S$ in the second order approximation substituting
the approximate expression for the evolution operator (\ref{eq:evol2}) into
Eq.\ (\ref{eq:supmatr}). Thus we have
\begin{equation}
S(\tau,t_0)=S^{(0)}(\tau)+S^{(1)}(\tau,t_0)+S^{(2)}(\tau,t_0)\label{eq:ap1}
\end{equation}
where $S^{(0)}(\tau)$ is the superoperator of the unperturbed measurement given
by Eq.\ (\ref{eq:s2}), $S^{(1)}(\tau,t_0)$ is the first order correction,
\begin{eqnarray}
S^{(1)}(\tau,t_0)_{p\alpha_3,r\alpha_4}^{n\alpha_1,m\alpha_2}
&=&\frac{1}{i\hbar}\delta_{rm}\delta(\alpha_4,\alpha_2)\exp(i
\omega_{r\alpha_4,p\alpha_3}\tau)\int_0^\tau dt V(t+t_0)_{p\alpha_3,n\alpha_1} \nonumber \\
&&\times\exp(i\omega_{p\alpha_3,n\alpha_1}t)F(\lambda(\omega_{rp}\tau
+\omega_{pn}t)) \nonumber \\
&&-\frac{1}{i\hbar}\delta_{pn}\delta(\alpha_3,\alpha_1)\exp(i
\omega_{r\alpha_4,p\alpha_3}\tau)\int_0^\tau dt V(t+t_0)_{m\alpha_2,r\alpha_4} \nonumber \\
&&\times\exp(i\omega_{m\alpha_2,r\alpha_4}t)F(\lambda(\omega_{rp}\tau
+\omega_{mr}t)),
\label{eq:ap2}
\end{eqnarray}
and $S^{(2)}(\tau,t_0)$ is the second order correction,
\begin{eqnarray}
S^{(2)}(\tau,t_0)_{p\alpha_3,r\alpha_4}^{n\alpha_1,m\alpha_2}
&=&\frac{1}{\hbar^2}\exp(i\omega_{r\alpha_4,p\alpha_3}\tau)\int_0^\tau
dt_1\int_0^\tau dt_2V(t_1+t_0)_{p\alpha_3,n\alpha_1}
V(t_2+t_0)_{m\alpha_2,r\alpha_4} \nonumber \\
&&\times F(\lambda(\omega_{rp}\tau+\omega_{pn}t_1+\omega_{mr}t_2))
\exp(i\omega_{p\alpha_3,n\alpha_1}t_1
+i\omega_{m\alpha_2,r\alpha_4}t_2) \nonumber \\
&&-\frac{1}{\hbar^2}\delta_{rm}\delta(\alpha_4,\alpha_2)\exp(i
\omega_{r\alpha_4,p\alpha_3}\tau)\sum_{s,\alpha} \nonumber \\
&&\int_0^\tau dt_1\int_0^{t_1}dt_2 V(t_1+t_0)_{p\alpha_3,s\alpha}
V(t_2+t_0)_{s\alpha,n\alpha_1}\nonumber \\
&&\times F(\lambda(\omega_{rp}\tau+\omega_{ps}t_1+\omega_{sn}t_2
))\exp(i\omega_{p\alpha_3,s\alpha}t_1
+i\omega_{s\alpha,n\alpha_1}t_2) \nonumber \\
&&-\frac{1}{\hbar^2}\delta_{pn}\delta(\alpha_3,\alpha_1)\exp(i
\omega_{r\alpha_4,p\alpha_3}\tau)\sum_{s,\alpha}\nonumber \\
&&\int_0^\tau dt_1\int_0^{t_1}dt_2 V(t_2+t_0)_{m\alpha_2,s\alpha}
V(t_1+t_0)_{s\alpha,r\alpha_4}\nonumber \\
&&\times F(\lambda(\omega_{rp}\tau+\omega_{sr}t_1+\omega_{ms}t_2
))\exp(i\omega_{s\alpha,r\alpha_4}t_1+i\omega_{m\alpha_2,s\alpha}t_2),
\label{eq:ap3}
\end{eqnarray}
where
\begin{equation}
\omega_{n\alpha_1,m\alpha_2}=\omega_{nm}+\frac{E_1(n,\alpha_1)-
E_1(m,\alpha_2)}{\hbar}.
\end{equation}

\end{document}